\documentclass[%
twocolumn%
,amsmath,amssymb,aps,prd,nobibnotes,
showpacs,
nofootinbib
,preprintnumbers
]{revtex4-1}
\usepackage{docs}
\expandafter\ifx\csname package@font\endcsname\relax\else
 \expandafter\expandafter
 \expandafter\usepackage
 \expandafter\expandafter
 \expandafter{\csname package@font\endcsname}%
\fi
\DeclareRobustCommand\substyle{\name@idx{document substyle}}%
\DeclareRobustCommand\classoption{\name@idx{document class option}}%
\DeclareRobustCommand\classname{\name@idx{document class}}%
\def\name@idx#1#2{%
 {\ttfamily#2}%
 \index{#2\space#1=\string\ttt{#2}\space#1}\index{#1>#2=\string\ttt{#2}}%
}%

\usepackage{graphicx}
\graphicspath{{./}}
\usepackage{bm,amssymb}
\usepackage{amsmath}
\usepackage{booktabs}
\usepackage{cases}
\usepackage{txfonts}
\usepackage{comment}
\usepackage{multirow}
\usepackage{color}
\newcommand{\Slash}[1]{{\ooalign{\hfil/\hfil\crcr$#1$}}}

\def\pbar{\overline{\psi}}

\def\Gs2{\Gamma^{(2)}_{k,\sigma}}
\def\Gp2{\Gamma^{(2)}_{k,\pi}}

\def\Jss{J_{k,\sigma\sigma}}
\def\Jpp{J_{k,\pi\pi}}
\def\Jsp{J_{k,\sigma\pi}}
\def\Jps{J_{k,\pi\sigma}}

\def\Is{I^{(2)}_{k,\sigma}}
\def\Ip{I^{(2)}_{k,\pi}}

\def\Jsqq{J^{(\sigma)}_{k,\bar{\psi}\psi}}
\def\Jpqq{J^{(\pi)}_{k,\pbar\psi}}

\def\Dk{\partial_{k}}

\def\RB{R^{B}_{k}}
\def\RF{R^{F}_{k}}

\def\ip0{\mathrm{i}p_{0}}
\def\p02{p_{0}^{2}}

\def\q2{\vec{q}^2}

\newcommand{\MeV}{\mathrm{MeV}}
\newcommand{\pab}{\ifmmode{p}\else{$p$}\fi}
\newcommand{\pfour}{\ifmmode{P}\else{$P$}\fi}
\newcommand{\qfour}{\ifmmode{Q}\else{$Q$}\fi}
\newcommand{\argmin}{\mathop{\rm{arg~min}}\limits}
\newcommand{\Ocp}{\ifmmode{\mathrm{O}(4)}\else{$\mathrm{O}(4)$}\fi}
\newcommand{\Zcp}{\ifmmode{\mathrm{Z}_{2}}\else{$\mathrm{Z}_{2}$}\fi}

\begin{document}
\title{
Tachyonic instability of the scalar mode prior to
QCD critical point \\
based on Functional renormalization-group method}
\author{Takeru Yokota}
\email{tyokota@ruby.scphys.kyoto-u.ac.jp}
\author{Teiji Kunihiro}
\email{kunihiro@ruby.scphys.kyoto-u.ac.jp}
\affiliation{Department of Physics, Faculty of Science, Kyoto University, Kyoto 606-8502, Japan}
\author{Kenji Morita}
\email{kmorita@yukawa.kyoto-u.ac.jp}
\affiliation{Yukawa Institute for Theoretical Physics, Kyoto University, Kyoto 606-8502, Japan \\
Institute of Theoretical Physics, University of Wroc{\l}aw, PL-50204, Wroc{\l}aw, Poland}

\preprint{KUNS-2693, YITP-17-74}

\begin{abstract}
We establish and elucidate the physical meaning of the appearance of an acausal
mode in the sigma mesonic channel, found in the previous work by the present authors,
when the system approaches the $\Zcp$ critical point.
The functional renormalization group method is applied to
the two--flavor quark--meson model with varying
current quark mass $m_q$ even away from the physical value
at which the pion mass is reproduced.
We first determine the whole phase structure in the
three-dimensional space $(T, \mu, m_q)$ consisting of temperature $T$,
quark chemical potential $\mu$ and $m_q$, with the tricritical point,
$\Ocp$ and $\Zcp$ critical lines being located;
they altogether make a wing-like shape
quite reminiscent of those known in the condensed matters with a tricritical point.
We then calculate the spectral functions $\rho_{\sigma, \pi}(\omega, p)$
in the scalar and pseudoscalar channel around the critical points.
We find that the sigma mesonic mode becomes tachyonic
with a superluminal velocity at finite momenta before the system
reaches the $\Zcp$ point from the lower density, even for $m_q$ smaller than the physical
value. One of the possible implications of the appearance of such a
tachyonic mode at finite momenta is that the assumed equilibrium state with
a uniform chiral condensate is unstable
toward a state with an inhomogeneous $\sigma$ condensate.
No such an anomalous behavior is found in the pseudoscalar channel.
We find that the $\sigma$-to-$2\sigma$ coupling due to finite $m_q$ play an essential role
for the drastic modification of the spectral function.
\end{abstract}

\maketitle

\section{Introduction}\label{Sec:Intro}

It is expected that hot and dense matter as described by the quantum
chromodynamics (QCD) shows a rich phase structure and the nature of the
phase transitions as well as the phase structure itself depends on the
external parameters such as temperature $T$, quark chemical potential
$\mu$, isospin chemical potential $\mu_I$ and so on. Furthermore, their
current quark-mass dependence adds interesting complications, which in
turn provide us with a theoretical clue to understand the mechanism of
the QCD phase transitions. For vanishing chemical potential, the
celebrated Columbia plot~\cite{PhysRevLett.65.2491} shows us how the
nature of the phase transitions may change along with that in the
current quark masses. At finite $\mu$ and $T$, the change of the phase
structure along with that of the current quark masses is expected as
follows. For simplicity, we shall take the two-flavor case composed of u
and d quarks and assume that $U_A(1)$ is kept broken at the chiral
phase transition~\footnote{There has been a debate on the fate of
U$_A(1)$ symmetry at high temperature since Ref.~\cite{Pisarski:1983ms},
though it is beyond our scope in this
paper. See, e.g.,\cite{Ishikawa:2017nwl, Tomiya:2016jwr} and references
therein for recent developments.}. In the
chiral limit, an $\Ocp$ critical line will exist in
the small $\mu$ region in accordance with the Columbia plot.
In the case of the  physical quark masses, it is considered that there
appears  a {\em first-order} phase transition line  in the large-$\mu$
and low-$T$ region while the phase change becomes crossover in the
small-$\mu$ and high-$T$ region
\cite{Asakawa:1989bq,Barducci:1989wi,Barducci:1989eu}.
The endpoint of the first-order phase transition line is of a {\em
second order} one and belongs to the same universality class as that of the
3D Ising model and is called a $\Zcp$ critical point: Indeed the current
quark mass $m_{q}$ plays the much the same role as the external magnetic
field in the Ising model, which can take positive and negative values.
The $\Zcp$ critical point moves as $m_q$ is varied and its trajectory
constitutes the $\Zcp$ critical lines in the three-dimensional
$(T,\mu,m_{q})$-space with fictitious negative $m_q$ axis being
included.
Thus the $\Ocp$ critical line bifurcates at some $\mu$ into the two
$\Zcp$ critical lines existing both in the positive and negative $m_{q}$
region resulting a wing-like structure \cite{kogut2010phases};
such a bifurcation point or the endpoint of the $\Ocp$ critical line
is called a tricritical point, and a similar structure appears in
various systems including metamagnets and
${}^{3}\mathrm{He}-{}^{4}\mathrm{He}$ mixtures
\cite{PhysRevB.7.545,gebhardt1980phasenubergange}.

One of the characteristics of any critical point 
is the existence of soft modes. It is also the case with the
$\Ocp$ and the $\Zcp$ critical points:
In the chiral limit, the sigma meson must be massless as are the three pions
at the $\Ocp$ critical point because of the $\Ocp$ symmetry.
Accordingly the soft modes composed of the pions and the sigma meson
form a quartet at the $\Ocp$ critical point.
In contrast, the nature of the soft modes on the $\Zcp$ critical line is
somewhat involved: The charge conjugation symmetry is broken due to a
finite $\mu$, in addition to the chiral symmetry owing to the finite
quark masses irrespective of current or dynamical ones. This gives rise
to a nonvanishing coupling between the fluctuations in Lorentz-scalar
and (the zero-th component of) vector channels, i.e., the baryon number
susceptibility \cite{Kunihiro:1991qu}, which should become divergent
on the $\Zcp$ critical line \cite{Kunihiro:2000ap}~\footnote{Precisely
speaking, the energy fluctuation can be also coupled to these
quantities, but the incorporation of the energy fluctuation is beyond
the scope of the present work.}.
Thus it is argued \cite{Fujii:2004jt,Son:2004iv} that the soft mode at
the $\Zcp$ critical point is mainly composed of
the particle--hole (p--h) modes describing density fluctuations or hydrodynamic
modes in the isoscalar channel while the mesonic mode mainly composed of
the $\sigma$ becomes a hard mode~\footnote{One notes that the spectral functions
$\rho_{\rm ph}(\omega, p)$ of the p--h modes have a support in the
space-like region ($\omega\,<\,p$) while the mesonic one in the
time-like region ($\omega\,>\,p$)}.
These analyses  were based on
 the random phase approximation (RPA) of
the Nambu--Jona-Lasinio model, the time-dependent Ginzburg--Landau theory
\cite{Fujii:2004jt} and the Langevin equation consisting solely of
the slow variables in the vicinity of the critical point \cite{Son:2004iv}.

Although the above analyses should be true at least in the very vicinity of
the $\Zcp$ critical point, the problem is how large is the critical region itself.
For a determination of the critical region, an analysis is necessary
which incorporates other  modes than the soft modes and their mutual
couplings systematically.
In the previous work \cite{Yokota:2016tip},
the present authors investigated the spectral properties of the
low-energy modes near a $\Zcp$ critical point using the functional
renormalization group (FRG) method, which is one of the non-perturbative
methods
of the field theory \cite{Berges:2000ew,Pawlowski:2005xe,Gies:2006wv}
and has been found to be useful in the description of the chiral phase
transition using effective chiral models
\cite{Jungnickel:1995fp,Braun:2003ii,Schaefer:2004en,Schaefer:2006ds,
Nakano:2009ps,Stokic:2009uv,Kamikado:2012cp,Aoki:2014ola,Morita:2014fda,
Morita:2014nra}.
In Ref.~\cite{Yokota:2016tip}, the quark--meson model was employed as the bare
action where the parameters including the current quark mass were
chosen so that the physical quantities such as the pion mass are
reproduced. It means that a special point on the $\Zcp$ critical line was
exclusively investigated.
The spectral function $\rho_{\sigma,\,\pi}(\omega, p)$ in the scalar and
pseudo-scalar channels were calculated, and the dispersion relations of
the mesonic and p--h modes were also extracted from the ridge of
$\rho_{\sigma,\,\pi}(\omega, p)$:
It was confirmed that the p--h phonon mode in the sigma channel certainly behaves
as the genuine soft mode
in the very vicinity of the $\Zcp$ critical point.
A surprise was that
the dispersion curve $\omega_{\sigma}(p)$
which is extracted as the ridge of $\rho_{\sigma}(\omega,p)$
 once sitting in the
time-like region penetrates into the space-like region with small and
vanishing momenta as the system approaches the critical point but still
before the phonon mode acquires the nature of the soft mode.
A notable point is that such an anomalous behavior
necessarily leads to
a superluminal group velocity of the mode, implying the appearance of the
tachyonic mode.
The system with $\Zcp$ critical point is characterized
by the explicit violation of the chiral symmetry and the charge conjugation symmetry
due to finite $m_q$ and $\mu$,
which in turn induce the couplings of the sigma with $2\sigma$ and
p--h excitation modes, respectively.
Thus the appearance and absence of the anomalous behavior of the sigma mode
mentioned above should be affected crucially
by the values of $m_q$ and $\mu$.

In the present work, we shall investigate the behavior of the low-energy
modes around {the $\Zcp$} critical point by varying the current quark
mass, and thereby establish and elucidate the physical meaning of the anomalous
behavior of the sigma-mesonic mode.
As in the previous work, we calculate the thermodynamic
quantities and the spectral functions in the mesonic channels using the
FRG method with the 2-flavor quark--meson model.
We first give a complete determination
of the phase structure in the $(T, \mu, m_q)$-space
with the tricritical point, $\Ocp$ and $\Zcp$ critical lines being located;
the resultant phase diagram
makes a wing-like shape which is
quite reminiscent of those known in the condensed matters with
 a tricritical point,
such as metamagnets and $^3$He--$^4$He mixtures.
It is then confirmed that the $\sigma$ meson is the soft mode of
the $\Ocp$ critical point in the chiral limit in the sense that the $\sigma$ mass
 tends to vanish as the system approaches the critical point.
Then it is shown for finite $m_q$  that
the $\sigma$ mesonic mode becomes
 superluminal  at finite momenta before the system
reaches the $\Zcp$ point from the lower density
even for $m_q$ finite but smaller than the physical
value. We argue that the appearance of such a
tachyonic mode with a superluminal velocity
at finite momenta may imply that the assumed equilibrium state with
 a uniform chiral condensate is unstable
for a phase transition
toward a state with an inhomogeneous $\sigma$ condensate.
We confirm that such a drastic change of the spectral
properties in the $\sigma$ channel is
attributed to the $\sigma$-to-$2\sigma$ coupling due to finite $m_q$.

The paper is organized as follows. In Sec.~\ref{Sec:Method}, we
 {give a brief description of the FRG method and the bare action.}
The parameter setting and the values of the current quark mass
used in our calculation are presented in Sec.~\ref{SubSec:Parameter}.
In Sec.~\ref{SubSec:CP},
 we give a complete phase diagram in the $(T, \mu, m_q)$-space
with the locations of the $\Ocp$  and the $\Zcp$ critical lines,
which makes a wing-like structure.
In Sec.~\ref{SubSec:Spectral}, we show the results of the spectral functions
in the mesonic channels near the critical points on the $\Ocp$ critical line
and the $\Zcp$ critical line and discuss the behavior of the low-energy modes
as the current quark mass is varied. Section~\ref{Sec:Summary} is
devoted to a summary.

\section{Formulation}\label{Sec:Method}

In this section, we briefly outline the method which was developed
in Refs.~\cite{Tripolt:2013jra,Tripolt:2014wra}
for the mesonic spectral functions based on the functional
renormalization group. More details can be found
in Ref.~\cite{Yokota:2016tip}.

In the functional renormalization group (FRG) method,
the effective average action (EAA) $\Gamma_{k}$,
which depends on the momentum scale $k$, is introduced.
The scale dependence of EAA follows the exact functional flow equation
\cite{Wetterich:1992yh}:
\begin{equation}
 \label{Eq:Wetterich}
 \partial_{k}\Gamma_{k}[\Phi]
 =
 \frac{1}{2}\mathrm{STr}\left[
 \frac{\partial_{k}R_{k}}{\Gamma^{(2)}_{k}[\Phi]+R_{k}}
 \right].
\end{equation}
Here, $\Phi$ represents all fields in the model
including bosons and fermions, and
$\Gamma^{(n)}_{k}[\Phi]\coloneqq \delta^{n}\Gamma_{k}/\delta \Phi^{n}$
($n$ is a natural number).
$R_{k}$ is a regulator function which  suppresses propagations of
lower momentum modes than  $k$.
The flow starts at large UV scale $k=\Lambda$, where EAA becomes the
bare action, and ends at  IR scale $k\rightarrow 0$, where all the
quantum fluctuations are incorporated and hence EAA becomes
the effective action. Physical quantities are derived from the solution
$\Gamma_{k\rightarrow 0}$ of Eq.~(\ref{Eq:Wetterich}).

To explore physics around the QCD critical point, we employ the two-flavor
quark--meson model which is a  chiral effective model of low-energy QCD.
The bare action at finite temperature ($T$) and finite quark chemical
potential ($\mu$) in the imaginary-time formalism is given by:
\begin{align}
 \label{Eq:Action}
 S_{\Lambda}\left[\pbar,\psi,\phi \right]
   & =
 \int_{0}^{\frac{1}{T}}\!\! d\tau
 \int d^{3}x
 \left\lbrace
 \pbar
 \left(\Slash{\partial}
 +g_{s}(\sigma+i\vec{\tau}\cdot\vec{\pi}\gamma_{5})
 -\mu \gamma_{0} \right)
 \psi\right. \nonumber\\
   & \left.+\frac{1}{2}(\partial_{\mu}\phi)^2
 +U_{\Lambda}(\phi^2)-c\sigma
 \right
 \rbrace.
\end{align}
Here, $\phi=(\sigma,\vec{\pi})$ denotes the chiral O$(4)$ multiplet
and $\sigma$ and $\vec{\pi}$ represent the $\sigma$ and $\pi$ meson fields,
respectively.
The quark field $\psi$ has the indices of the flavor $N_{f}=2$
and the color $N_{c}=3$.
The quarks are coupled with the mesonic fields through the Yukawa
coupling whose strength is represented as $g_{s}$.
$U_{\Lambda}(\phi^{2})$ is the potential term of the mesons.
The last term $-c\sigma$ corresponds to the nonzero current quark mass
thus explicitly breaks the $N_f=2$ chiral symmetry when $c\neq 0$.

Our purpose is to calculate the mesonic spectral functions
$\rho_{\sigma(\pi)}$.
which can be obtained from the imaginary parts of
the retarded Green's functions $G^{R}_{\sigma(\pi)}$:
\begin{equation}
 \label{Eq:Spectral}
 \rho_{\sigma (\pi)}(\omega, \vec{p})
 =
 -\frac{1}{\pi}\mathrm{Im}G^{R}_{\sigma (\pi)}(\omega,\vec{p}).
\end{equation}
In the imaginary-time formalism, the retarded Green's functions
can be derived from the analytic continuations of
the temperature Green's functions \cite{Abrikosov:107441}.

We define the scale-dependent temperature Green's functions
$\mathcal{G}_{k,\sigma(\pi^a)}(P)$ for the sigma and the pion with isovector
index $a$ as follows:
\begin{align}
 \left. \frac{\delta^{2} \Gamma_{k}}{\delta \sigma(P) \delta \sigma(Q)}
 \right|_{\Phi=\Phi_0}
   & =
 (2\pi)^{4}\delta(P+Q)\mathcal{G}_{k,\sigma}^{-1}(P), \\
 \left. \frac{\delta^{2} \Gamma_{k}}{\delta \pi^{a}(P) \delta \pi^{a}(Q)}
 \right|_{\Phi=\Phi_0}
   & =
 (2\pi)^{4}\delta(P+Q)\mathcal{G}_{k,\pi^a}^{-1}(P).
\end{align}
Here the four-momenta $P$ and $Q$ have a Matsubara frequency
and the space momentum $\vec{P}$ and $\vec{Q}$ in the
temporal and spatial components, respectively.
The field variables are taken to be their ground state
expectation values.
$\mathcal{G}_{k,\sigma}(P)$ and $\mathcal{G}_{k,\pi^a}(P)$ converge on
the temperature Green's functions for the respective mesons as $k\rightarrow 0$.
The flow equations for $\mathcal{G}_{k,\sigma(\pi)}(P)$
\footnote{
From now on we consider only isospin symmetric case and
abbreviate the isospin index for pions.}
is obtained from the second derivatives of Eq.~(\ref{Eq:Wetterich})
with respect to the meson fields:
\begin{align}
 \label{Eq:Flowsigma}
 (2\pi)^{4}\delta^{(4)}(P+Q)\partial_{k}\mathcal{G}_{k,\sigma}^{-1}(P)
   & =
 \frac{1}{2}\left. \frac{\delta^{2}}{\delta \sigma(P) \delta \sigma(Q)}
 \left[
 \frac{\partial_{k}R_{k}}{\Gamma^{(2)}_{k}+R_{k}}
 \right]
 \right|_{\Phi=\Phi_0}, \\
 \label{Eq:Flowpi}
 (2\pi)^{4}\delta^{(4)}(P+Q)\partial_{k}\mathcal{G}_{k,\pi^a}^{-1}(P)
   & =
 \frac{1}{2}\left. \frac{\delta^{2}}{\delta \pi^{a}(P) \delta \pi^{a}(Q)}
 \left[
 \frac{\partial_{k}R_{k}}{\Gamma^{(2)}_{k}+R_{k}}
 \right]
 \right|_{\Phi=\Phi_0}.
\end{align}

Equations \eqref{Eq:Flowsigma} and \eqref{Eq:Flowpi} contain
the third and fourth derivatives of EAA in the right-hand side.
The flow equation for $\Gamma_{k}^{(n)}$ contains
$\Gamma_{k}^{(n+1)}$ and $\Gamma_{k}^{(n+2)}$ leading to an infinite
hierarchy of a coupled equation, and thus it is difficult to
obtain the exact solution of Eq.~(\ref{Eq:Wetterich}) in general.
Therefore some approximations are needed
to reduce the infinitely coupled
differential equations to a solvable form.
In the present work, we are interested in the low-energy modes around
the critical point, which in turn play the dominant role for the
determination of the phase structure around there.
Thus
we take so-called the local potential approximations where only the
leading order in the derivative expansion of the meson fields is
considered and the wave-function renormalization is neglected,
and assume the following form for the EAA in the present work
\cite{Schaefer:2004en}:

\begin{widetext}

 \begin{align}
  \label{Eq:Truncation}
  \Gamma_{k}\left[\pbar,\psi,\phi \right]
  =
  \int_{0}^{\frac{1}{T}}\!\! d\tau
  \int d^{3}x
  \left\lbrace
  \pbar
  \left(\Slash{\partial}
  +g_{s}(\sigma+i\vec{\tau}\cdot\vec{\pi}\gamma_{5})
  -\mu \gamma_{0} \right)
  \psi
  +\frac{1}{2}(\partial_{\mu}\phi)^2
  +U_{k}(\phi^2)-c\sigma
  \right
  \rbrace ,
 \end{align}
 where $U_{k}(\phi^{2})$ is the scale-dependent potential (effective potential)
 for the mesons.

 The analytic continuation of the Green's function
 can be performed in the flow equation by making use of three-dimensional
 regulators which do not depend on frequency
 \cite{Kamikado:2013sia,Tripolt:2013jra,Tripolt:2014wra,Floerchinger:2011sc}.
 We adopt the three-dimensional forms of the so-called optimized
 regulators, proposed by Litim \cite{Litim:2001up}:
 \begin{align}
  \label{Eq:Rb}
  \RB(\vec{q})
    & =
  (k^2-\vec{q}^2)\theta(k^2-\vec{q}^2), \\
  \label{Eq:Rf}
  \RF(\vec{q})
    & =
  i\Slash{\vec{q}}\left(\sqrt{\frac{k^2}{\vec{q}^2}}-1\right)
  \theta(k^2-\vec{q}^2).
 \end{align}
 Here, $\RB(\vec{q})$ and $\RF(\vec{q})$ are the regulators for bosons
 and  fermions, respectively.

 With the above regulators and assumptions of the uniform condensation of
 $\sigma$ field and no $\pi$ condensate in the ground state,
 $\sigma_0 =\langle \sigma \rangle$ and $\langle \vec{\pi} \rangle=0$,
 the flow equation for $U_{k}$ reads \cite{Schaefer:2004en}:
 \begin{align}
  \label{Eq:Ukflow}
  \partial_{k}U_k (\sigma^{2})
  =\frac{k^{4}}{12\pi^2}
  \left[
  -2N_f N_c
  \left[
  \frac{1}{E_{\psi}(\sigma)}
  \tanh\frac{E_{\psi}(\sigma)+\mu}{2T}
  +
  \frac{1}{E_{\psi}(\sigma)}
  \tanh\frac{E_{\psi}(\sigma)-\mu}{2T}
  \right]
  +
  \frac{1}{E_{\sigma}(\sigma)}
  \coth\frac{E_{\sigma}(\sigma)}{2T}
  +
  \frac{3}{E_\pi(\sigma)}
  \coth\frac{E_\pi(\sigma)}{2T}
  \right],
 \end{align}
 where
 \begin{equation}
  E_{a}(\sigma)\coloneqq \sqrt{k^{2}+M_{a}^{2}(\sigma)},
 \end{equation}
 for $a=\psi,\ \sigma,\ \pi$, and $M_{a}(\sigma)$ is defined as follows:
 \begin{align}
  M_{\psi}(\sigma)
    & \coloneqq
  g_{s}\sigma, \\
  M_{\sigma}^{2}(\sigma)
    & \coloneqq
  \frac{\partial^{2}U_{k}}{\partial\sigma^{2}},\label{eq:m_scr-sigma} \\
  M_{\pi}^{2}(\sigma)
    & \coloneqq
  \frac{1}{\sigma}\frac{\partial U_{k}}{\partial\sigma}.\label{eq:m_scr-pi}
 \end{align}
 After solving the flow equation, the $\sigma$ condensate
 $\sigma_{0}$ is determined from the minimum of the effective potential
 $U_{k\rightarrow 0}$:
 \begin{equation}
  \label{Eq:Sigma0}
  \sigma_{0}=\argmin_{\sigma}(U_{k\rightarrow 0}(\sigma^2)-c\sigma).
 \end{equation}

 The flow equations of the temperature Green's function
 (\ref{Eq:Flowsigma}) and (\ref{Eq:Flowpi}) can be evaluated
 with $U_{k}$ and $\sigma_{0}$.
 The analytic continuation of
 $\mathcal{G}_{k,\sigma(\pi)}(P)$ by the
 replacement $iP^{0}\rightarrow \omega + i\epsilon$ for the Matsubara frequency
 with $\epsilon$ being
 a positive infinitesimal number
 give the scale-dependent retarded Green's functions
 $G^{R}_{k,\sigma(\pi)}(\omega, \vec{p})$, which are to converge on
 the retarded Green's functions for the mesons as $k\rightarrow 0$:
 The flow equations for $G^{R}_{k,\sigma(\pi)}(\omega, \vec{p})$ now read
 \cite{Tripolt:2013jra,Tripolt:2014wra}:
 \begin{align}
  \label{Eq:Gsflow}
  \Dk G^{R}_{k,\sigma}(\omega, \vec{p})^{-1}
  = &
  \left. \left[
  \Jss(\pfour)(\Gamma^{(0,3)}_{k,\sigma\sigma\sigma})^{2}
  +3\Jpp(\pfour)(\Gamma^{(0,3)}_{k,\sigma\pi\pi})^{2}
  -\frac{1}{2}\Is \Gamma^{(0,4)}_{k,\sigma\sigma\sigma\sigma}
  -\frac{3}{2}\Ip \Gamma^{(0,4)}_{k,\sigma\sigma\pi\pi}
  -2N_{c}N_{f}\Jsqq(\pfour)
  \right]\right|_{iP^{0}\rightarrow \omega + i\epsilon}, \\
  \label{Eq:Gpflow}
  \Dk G^{R}_{k,\pi}(\omega, \vec{p})^{-1}
  = &
  \left.\left[
  \Jsp(\pfour)(\Gamma^{(0,3)}_{k,\sigma\pi\pi})^{2}+\Jps(\pfour)
  (\Gamma^{(0,3)}_{k,\sigma\pi\pi})^{2}
  -\frac{1}{2}\Is\Gamma^{(0,4)}_{k,\sigma\sigma\pi\pi}
  -\frac{5}{2}\Ip\Gamma^{(0,4)}_{k,\pi\pi\tilde{\pi}\tilde{\pi}}
  -2N_{c}N_{f}\Jpqq(\pfour)
  \right]\right|_{iP^{0}\rightarrow \omega + i\epsilon}.
 \end{align}
 Here, $\tilde{\pi}$ denotes a pion field with a isovector component
 different from $\pi$.
 $J_{k,\alpha\beta}(P)$,\, $I^{(2)}_{k,\alpha}$ and
 $J_{k,\pbar \psi}^{(\alpha)}(P)$\, $(\alpha,\beta=\sigma,\pi)$ are defined
 as
 \begin{align}
  \label{Eq:Jab}
  J_{k,\alpha\beta}(\pfour)
    & \coloneqq
  T\sum_{Q^{0}}\int \frac{d^{3}\vec{q}}{(2\pi)^{3}} \partial_{k}
  R_{k}^{B}(\vec{q}) G^{B}_{k,\alpha}(P)^{2} G^{B}_{k,\beta}(Q-P),
  \\
  \label{Eq:Ia}
  I^{(2)}_{k,\alpha}
    & \coloneqq
  T\sum_{Q^{0}}\int \frac{d^{3}\vec{q}}{(2\pi)^{3}} \partial_{k}
  R_{k}^{B}(\vec{q})G^{B}_{k,\alpha}(Q)^{2},
  \\
  \label{Eq:Jqq}
  J_{k,\pbar \psi}^{(\alpha)}(\pfour)
    & \coloneqq
  T\sum_{Q^{0}}\int \frac{d^{3}\vec{q}}{(2\pi)^{3}} \mathrm{tr}
  \left[
  \Gamma^{(2,1)}_{\pbar\psi\alpha}
  G^{F}_{k,\pbar \psi}(Q)\partial_{k}R_{k}^{F}(\vec{q})
  G^{F}_{k,\pbar \psi}(Q)\Gamma^{(2,1)}_{\pbar\psi\alpha}
  G^{F}_{k,\pbar \psi}(Q-P)
  \right],
 \end{align}
 where $G^{B}_{k,\alpha}(\qfour)$ and $G^{F}_{k,\pbar \psi}(\qfour)$
 are defined as
 \begin{align}
  G^{B}_{k,\alpha}(\qfour)
    & \coloneqq
  \left[\qfour^{2}+M_{\alpha}^{2}(\sigma_{0})+R_{k}^{B}(\vec{q}) \right]^{-1},
  \label{Eq:GB}
  \\
  G^{F}_{k,\pbar \psi}(\qfour)
    & \coloneqq
  \left[\Slash{\qfour}-\mu \gamma_{0}+
  M_{\psi}(\sigma_{0})+R^{F}_{k}(\vec{q})\right]^{-1}.
  \label{Eq:GF}
 \end{align}
 We refer to Appendix A of Ref.~\cite{Yokota:2016tip} for more practical
 expressions of Eqs.~\eqref{Eq:Jab}-\eqref{Eq:Jqq} after the Matsubara
 summation.
 Note, that in our truncation, the meson propagator \eqref{Eq:GB} in the
 flow equation depends on the scale-dependent
 meson \emph{screening} mass
 \eqref{eq:m_scr-sigma}--\eqref{eq:m_scr-pi}.
 We will discuss some features of the resultant spectral function owing
 to this approximation in Sec.~\ref{Sec:discussion}.

 The three- and four-point functions
 $\Gamma^{(2,1)}_{\pbar\psi\phi_{i}}$,
 $\Gamma^{(0,3)}_{k,\phi_{i}\phi_{j}\phi_{l}}$, and
 $\Gamma^{(0,4)}_{k,\phi_{i}\phi_{j}\phi_{l}\phi_{m}}$
 are defined as
 \begin{align}
  \left.
  \frac{\delta}{\delta \phi_{i} (P_{1})}
  \frac{\overset{\rightarrow}{\delta}}{\delta \pbar (P_{2})}
  \Gamma_{k}
  \frac{\overset{\leftarrow}{\delta}}{\delta \psi (P_{3})}
  \right|_{\Phi=\Phi_0}
    & =
  (2\pi)^{4}\delta^{(4)}(P_{1}+P_{2}+P_{3})
  \Gamma^{(2,1)}_{\pbar\psi\phi_{i}},
  \\
  \left.
  \frac{\delta^{3} \Gamma_{k}}{
  \delta \phi_{i} (P_{1})
  \delta \phi_{j} (P_{2})
  \delta \phi_{l} (P_{3})}
  \right|_{\Phi=\Phi_0}
    & =
  (2\pi)^{4}\delta^{(4)}(P_{1}+P_{2}+P_{3})
  \Gamma^{(0,3)}_{k,\phi_{i}\phi_{j}\phi_{l}}, \\
  \left.
  \frac{\delta^{4} \Gamma_{k}}{
  \delta \phi_{i} (P_{1})
  \delta \phi_{j} (P_{2})
  \delta \phi_{l} (P_{3})
  \delta \phi_{m} (P_{4})}
  \right|_{\Phi=\Phi_0}
    & =
  (2\pi)^{4}\delta^{(4)}(P_{1}+P_{2}+P_{3}+P_{4})
  \Gamma^{(0,4)}_{k,\phi_{i}\phi_{j}\phi_{l}\phi_{m}}.
 \end{align}
\end{widetext}

The initial conditions of the flow equations are specified at the UV
scale $k=\Lambda$ from the the bare action \eqref{Eq:Action}:
\begin{align}
 \label{initial}
 U_{\Lambda}(\phi^2)
   & =
 \frac{1}{2}m_{\Lambda}^2\phi^2+\frac{1}{4}\lambda_{\Lambda}(\phi^2)^2,\\
 G^{R}_{\Lambda,\sigma(\pi)}(\omega,\vec{p})^{-1}
   & =
 \omega^{2}-\vec{p}^{2}
 -\left. M_{\sigma(\pi)}^{2}(\sigma_{0}^{2})\right|_{k=\Lambda}.
\end{align}

Before closing the section, we
briefly describe our numerical implementations.
We use the grid method to solve Eq.~(\ref{Eq:Ukflow}). Thus higher
powers of $\phi^2$ in $U_k$ are automatically incorporated.
As pointed out in Ref.~\cite{Yokota:2016tip},
the grid intervals $\Delta \sigma$ and $\Delta t$ for $\sigma$ and
$t=\exp(k)$, respectively,
should satisfy a condition to maintain numerical instability in the
evolution of the flow.
We fix $\Delta \sigma=0.32~\MeV$ and set $\Delta t$ so that
the condition is satisfied.
We introduce an IR cutoff $k=k_{\mathrm{IR}}$ at which the evolution is stopped
because the stability condition makes the calculation time-consuming
beyond that scale.
In practice, $k_{\mathrm{IR}}$ is set to be smaller than $3~\MeV$ in our
calculation.
The calculation is expected to be reliable for the analysis of modes
with greater momentum scale than $k_{\mathrm{IR}}$.
The positive infinitesimal $\epsilon$ in the replacement of the
frequency is set to $\epsilon=1~\MeV$.

\section{Numerical results and discussions}\label{Sec:Results}

\begin{table*}[!tb]
 \centering
 \caption{Parameter set for the physical vacuum.}
 \label{Tab:Parameters}
 \begin{tabular}{ccccc|cccc}
  \hline
  $\Lambda$   & $m_{\Lambda}/\Lambda$ & $\lambda_{\Lambda}$ &
  $c/\Lambda^{3}$ & $g_{s}$ &
  $\sigma_0$  & $m_{\text{con}}$      & $m_\pi$             & $m_\sigma$
  \\ \hline
  $1000~\MeV$ & 0.794                 & 2.00                &
  0.00175     & 3.2                   & $93~\MeV$           &
  $286~\MeV$ & $137~\MeV$ & $496~\MeV$
  \\ \hline
 \end{tabular}
\end{table*}

In this section, we first show how the parameters in our model
are determined
with a special attention to the current quark mass
since our EAA does not explicitly contain the current quark mass term;
the explicit breaking of chiral symmetry is given by the last term
$-c\sigma$ in the action.
Then we proceed to show the numerical results
on the phase diagram and the spectral properties
of the low-energy modes with the current quark mass being varied.

{
 \begin{table*}[!bt]
  \centering
  \caption{Explicit breaking parameter $c$ explored in this work and
   corresponding particle masses from Eq.~\eqref{eq:mq}
   \label{Tab:cValues}}
  \begin{tabular}{c|cccccccccc}
   \hline
   $c/\Lambda^{3}$   & 0   & 0.00025 & 0.0005 & 0.00075 & 0.001 & 0.00125
   & 0.0015 & 0.00175 & 0.002 & 0.003 \\ \hline
   $m_{q}~[\MeV]$    & 0   & 1.27    & 2.54   & 3.81    & 5.08  & 6.35
   & 7.61   & 8.88    & 10.15 & 15.23 \\ \hline
   $m_\pi~[\MeV]$    & 0   & 54      & 76     & 92      & 105   & 117
   & 127    & 137     & 146   & 175   \\ \hline
   $m_\sigma~[\MeV]$ & 111 & 336     & 390    & 423     & 448   & 467
   & 483    & 496     & 508   & 546   \\ \hline
  \end{tabular}
 \end{table*}
}

\subsection{Parameter setting}
\label{SubSec:Parameter}

We first show the physical parameters which reproduce
the empirical value of the pion mass
in the vacuum \cite{Tripolt:2013jra,Tripolt:2014wra,Yokota:2016tip}.
The parameter values
and resultant values of
some observables in the vacuum are shown in Table~\ref{Tab:Parameters},
where $m_{\mathrm{con}}$ and $m_{\sigma (\pi)}$
are defined as follows:
\begin{align}
 m_{\mathrm{con}}
   & =\left. M_{\psi}(\sigma_{0})\right|_{k\rightarrow 0},         \\
 m_{\sigma (\pi)}
   & =\left. M_{\sigma (\pi)}(\sigma_{0})\right|_{k\rightarrow 0}.
\end{align}

For analyzing the cases with various current quark masses,
we vary the parameter $c$ which represents the effect of explicit breaking of
the chiral symmetry in our model
and can be related to the current quark mass $m_q$, as shown below:
Other parameters  such as $\Lambda$, $m_{\Lambda}$, $\lambda_{\Lambda}$
and $g_s$ are fixed to the values listed in Table~\ref{Tab:Parameters}.
The classical relation between $\sigma$ and $\pbar \psi$
is given by the variation of Eq.~(\ref{Eq:Action})
with respect to $\sigma$.
Neglecting the kinematic term and the high-order terms of $\sigma$,
one finds
\begin{equation}
 \sigma \sim -\frac{g_{s}}{m_{\Lambda}^{2}}\pbar\psi.\label{eq:sigma-psibarpsi}
\end{equation}
Substituting Eq.~\eqref{eq:sigma-psibarpsi} into Eq.~(\ref{Eq:Action}),
the current quark mass can be read off from the quark mass term
$-(cg_{s}/m_{\Lambda}^{2})\pbar\psi$ as
\begin{equation}
 m_{q}\sim \frac{cg_{s}}{m_{\Lambda}^{2}}.\label{eq:mq}
\end{equation}
The values of $c$ and the corresponding $m_{q}$ as well as
the meson masses are listed in Table~\ref{Tab:cValues}.
We note that the Gell-Mann--Oakes--Renner (GOR) relation gives the almost
the same quark masses.

{
 \begin{table*}[!bt]
  \centering
  \caption{
   The position of the {\Zcp} critical point $( T_{c}(m_{q}),~\mu_{c}(m_{q}))$
   for each $m_{q}$.
   \label{Tab:CP}}
  \begin{tabular}{c|ccccccccc}
   \hline
   $m_{q}~[\MeV]$
     & 1.27 & 2.54 & 3.81 & 5.08 & 6.35
   & 7.61 & 8.88 & 10.15 & 15.23
   \\ \hline
   $T_{c}(m_{q})~[\MeV]$
     & 8.6  & 7.7  & 6.9  & 6.3  & 5.9
   & 5.5 & 5.1 & 4.8 & 4.0
   \\ \hline
   $\mu_{c}(m_{q})~[\MeV]$
   & 264.3059 & 268.8192 & 272.8095 & 276.5925
   & 280.2998 & 283.7038 & 286.8499
   & 289.9452 & 301.4115  \\ \hline
  \end{tabular}
 \end{table*}
}

\subsection{Positions of critical points}
\label{SubSec:CP}

In this subsection, we determine the position of the critical point
in the $(T,~\mu)$-plane for each current mass.
We remark that for determination of the phase structure
with the critical point being identified,
we need to calculate the $sigma$ condensate $\sigma_0=\langle \sigma \rangle$
for a wide region of
the $(T,~\mu)$-plane.

{
 \begin{figure}[!t]
  \centering\includegraphics[width=0.95\columnwidth]{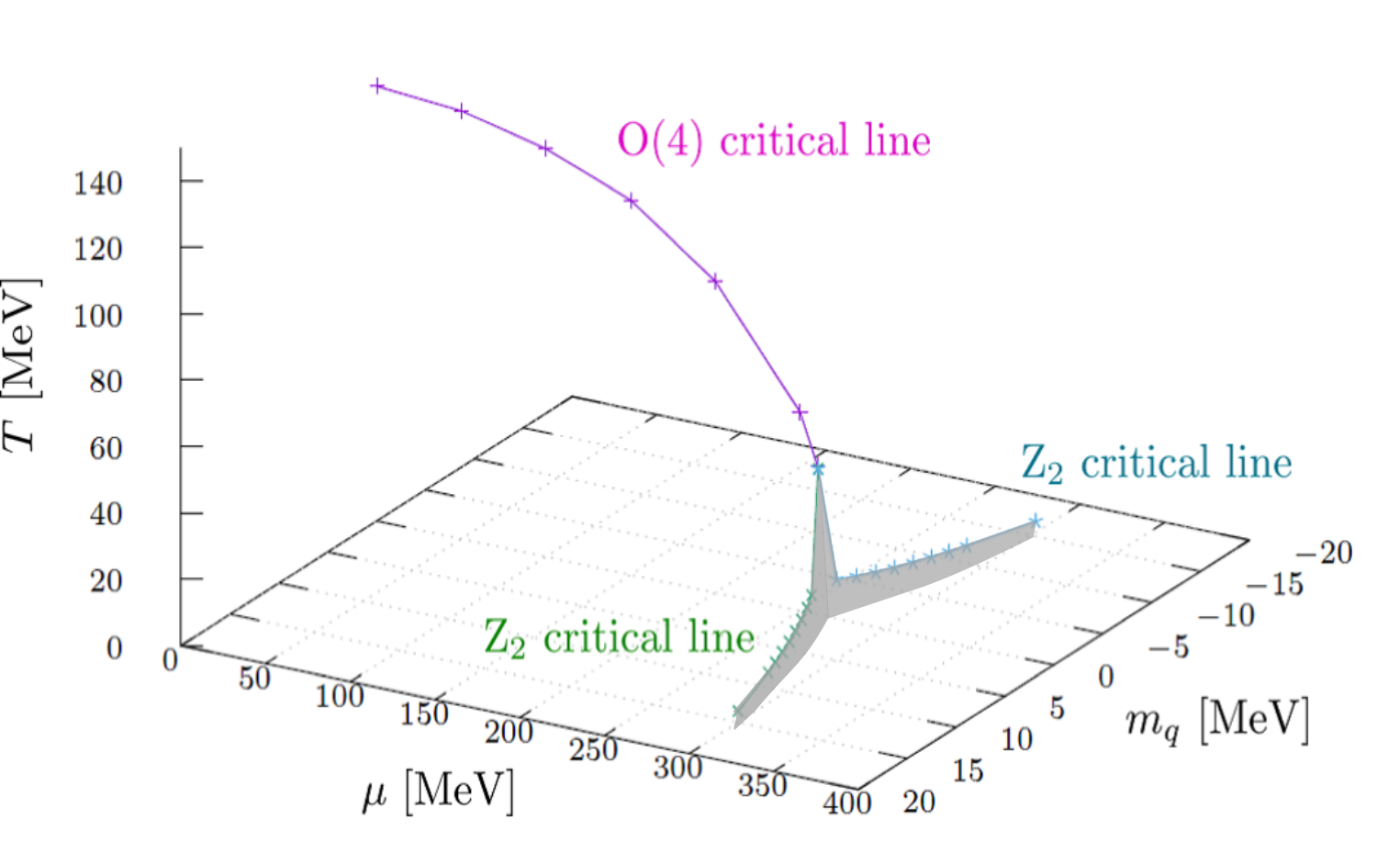}
  \caption{
   The calculated phase structure in the $(T$,\ $\mu$,\ $m_q)$-space
   with the $\Ocp$ and $\Zcp$ critical lines being identified.
   The shaded area indicates that the phase transition is first order there.
   The phase diagram for the fictitious case with $m_q < 0$ is also shown:
   The phase structure is symmetric with respect to the $m_{q}=0$ plane
   because the theory is invariant under the transformation
   $m_{q}\rightarrow -m_{q}$ and global phase transformations of fields.
   As mentioned in Sec.~\ref{Sec:Intro}, such a wing-like structure
   is common to various systems which have a tricritical point
   \cite{Fujii:2004jt,kogut2010phases,PhysRevB.7.545,
   gebhardt1980phasenubergange}.
   \label{Fig:CPlocation}}
 \end{figure}
}

\subsubsection{$\Ocp$ critical line; chiral limit}

In the case of the chiral limit,
the chiral transition is of a second order at finite temperature for
vanishing or small $\mu$. Thus the critical line, i.e.,
an $\Ocp$ critical line exists in the high-temperature
and low-density region starting from a point $(T\not=0,~\mu=0)$.
The $\Ocp$ critical line terminates at the tricritical point
where it is connected to the first-order
phase transition line in the high-density and low-temperature region.
To specify the phase boundary, we search the points where $\sigma_{0}$
changes from 0 to a nonzero value in the $(T,~\mu)$-plane.
Then, we examine whether a gap in $\sigma_0$ appears or not to
distinguish the order of the phase transition.
Around the tricritical point, we calculate $\sigma_{0}$ with changing
the chemical potential with an interval of $0.0001~\MeV$ for each
temperature with $1~\MeV$ intervals. A gap of $\sigma_{0}$ is found to
appear at $T=44~\MeV$, but does not appear at $T=45~\MeV$.
Thus we conclude that a  first-order phase transition at $T=44~\MeV$
and the tricritical point  between $44~\MeV$ and $45~\MeV$.
The location of the $\Ocp$ critical line is depicted
in Fig.~\ref{Fig:CPlocation}
in which the $\Zcp$ lines for finite $m_q$ are also shown; see below.

\subsubsection{$\Zcp$ critical line; $m_q\not=0$}

For
$m_q\not=0$, the phase transition becomes of crossover
in the lower-chemical potential region while the first-order phase
transition line persists in the high chemical potential region.
At the endpoint of the first-order phase transition line,
the phase transition is of  second order,
and accordingly the screening mass $m_{\sigma}^{2}$,
which is the inverse of the chiral susceptibility,
vanishes. This critical point is called
the $\Zcp$ critical point.
To locate the $\Zcp$ critical point $(T_c(m_q),\ \mu_{c}(m_q))$, we search
the point where  $m_{\sigma}$ takes the smallest value by varying $T$ and
$\mu$ with an interval of $\Delta T=0.1~\MeV$
and $\Delta \mu=0.0001~\MeV$, respectively.
We also confirmed that the phase change is of crossover at temperatures
above the $\Zcp$ critical temperature $T_c(m_q)$ and
of first-order at temperatures below $T_c(m_q)$,
by investigating the gap of $\sigma_{0}$.
The location of the $\Zcp$ critical point for each $m_{q}$ is depicted
in Fig.~\ref{Fig:CPlocation}, and the sets of values of
$(T_{c}(m_{q}),~\mu_{c}(m_{q}))$ for each $m_{q}$
are shown in Table~\ref{Tab:CP}.
One finds that as $m_{q}$ becomes larger, the critical point shifts to
the low-temperature and high chemical potential region.

\subsection{Behavior of low-energy modes around the critical points}
\label{SubSec:Spectral}

{
 \begin{figure}[!b]
  \centering\includegraphics[width=0.5\columnwidth]{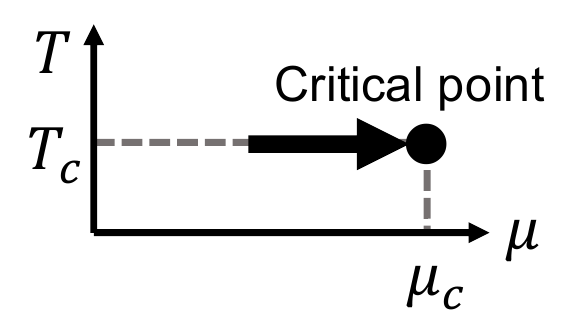}
  \caption{
   A schematic view on how to approach a critical point in our calculations.
   The temperature is fixed to the critical temperature $T_{c}(m_q)$
   and the chemical potential $\mu$ approaches the critical chemical potential
   $\mu_{c}(m_q)$ from the region $\mu < \mu_{c}(m_q)$.
   \label{Fig:Direction}}
 \end{figure}
}

In this section, we show the results of the spectral functions
$\rho_{\sigma,\pi}$ in the mesonic channels, almost exclusively focusing
on that in the sigma channel $\rho_{\sigma}$;
we discuss the behavior of low-energy modes
around the critical points when
the current quark mass is varied.

Before entering into the results, we list up the possible physical
processes which may give rise to a peak or bump in the  spectral
functions $\rho_{\sigma,\pi}$ as preliminaries \cite{Yokota:2016tip}.
In addition to one-particle modes of the mesons, decay and absorption
processes of the excitation modes cause a peak or bump in the spectral
functions.
The decay and absorption processes together with the respective
kinematic conditions in the sigma and pion channels are shown in
Table~\ref{Tab:Modes}.
In particular, the process $\sigma^{\ast}\psi \rightarrow \psi$
denotes a collisional process of the virtual $\sigma$ state with quarks,
which leads to (quark) particle--hole (p--h) excitations of a phonon type.
This is the so-called  Landau damping and is known to
play an essential role to make  the specific
soft mode at the {\Zcp} critical point \cite{Fujii:2004jt}.

{
 \begin{table}[]
  \centering
  \caption{
   Decay and absorption processes in the sigma and pion channels and
   their kinematic constraints.
   $\sigma^{\ast}$ and $\pi^{\ast}$ represent virtual particles
   in the sigma  and pion channels, respectively.
   $\omega$ and $\vec{p}$ are the energy and momentum of the virtual
   particles. $\alpha$ represents $\sigma$ and $\pi$.
   \label{Tab:Modes}}
  \begin{tabular}{ccc}
   \hline
   Channel
     & Process
     & Constraint
   \\ \hline
   \multirow{4}{*}{Sigma meson}
     & $\sigma^{\ast}\rightarrow \alpha\alpha$
     & $\omega \geq \sqrt{\vec{p}^{2}+(2m_{\alpha})^{2}}$
   \\ \cline{2-3}
     & $\sigma^{\ast}\rightarrow \pbar\psi$
     & $\omega \geq \sqrt{\vec{p}^{2}+(2m_{\mathrm{con}})^{2}}$
   \\ \cline{2-3}
     & $\sigma^{\ast}\alpha \rightarrow \alpha$
     & $\omega \leq \left| \vec{p}\right|$
   \\ \cline{2-3}
     & $\sigma^{\ast}\psi \rightarrow \psi$
     & $\omega \leq \left| \vec{p}\right|$
   \\ \hline
   \multirow{5}{*}{Pion}
     & $\pi^{\ast}\rightarrow \sigma\pi$
     & $\omega \geq \sqrt{\vec{p}^{2}+(m_{\sigma}+m_{\pi})^{2}}$
   \\ \cline{2-3}
     & $\pi^{\ast}\rightarrow \pbar\psi$
     & $\omega \geq \sqrt{\vec{p}^{2}+(2m_{\mathrm{con}})^{2}}$
   \\ \cline{2-3}
     & $\pi^{\ast}\sigma\rightarrow \pi$
     & $\omega \leq\sqrt{\vec{p}^{2}+(m_{\pi}-m_{\sigma})^{2}
   \theta(m_{\pi}-m_{\sigma})}$
   \\ \cline{2-3}
     & $\pi^{\ast}\pi \rightarrow \sigma$
     & $\omega \leq \sqrt{\vec{p}^{2}+(m_{\sigma}-m_{\pi})^{2}
   \theta(m_{\sigma}-m_{\pi})}$
   \\ \cline{2-3}
     & $\pi^{\ast}\psi \rightarrow \psi$
     & $\omega \leq \left| \vec{p}\right|$
   \\ \hline
  \end{tabular}
 \end{table}
}

Now that we have finished the preliminaries, we show how the properties
of the low-energy excitations in the scalar channel as extracted from
the spectral function as
$\mu$ approaches the critical
chemical potential $\mu_c(m_q)$ from below at fixed temperature
$T=T_c(m_q)$ for each current quark mass; see Fig.~\ref{Fig:Direction}.
A remark is in order here:
For
$m_q\not= 0$,
$M_{\sigma}^{2}(\sigma_{0})$ tends to take a negative value which causes
numerical instability at some small $k$ during the flow when
$\mu > \mu_c$,
and we restrict ourselves to the cases with $\mu$ smaller than
the critical value.

We consider the spectral function not only vanishing momentum but also
finite $\vec{p}$. Note that spectral functions depend only on
$p = \left|\vec{p}\right|$, since the system is isotropic.

\subsubsection{Appearance of tachyonic mode at physical quark mass}

{
 \begin{figure*}[!htb]
  \centering\includegraphics[width=1.7\columnwidth]{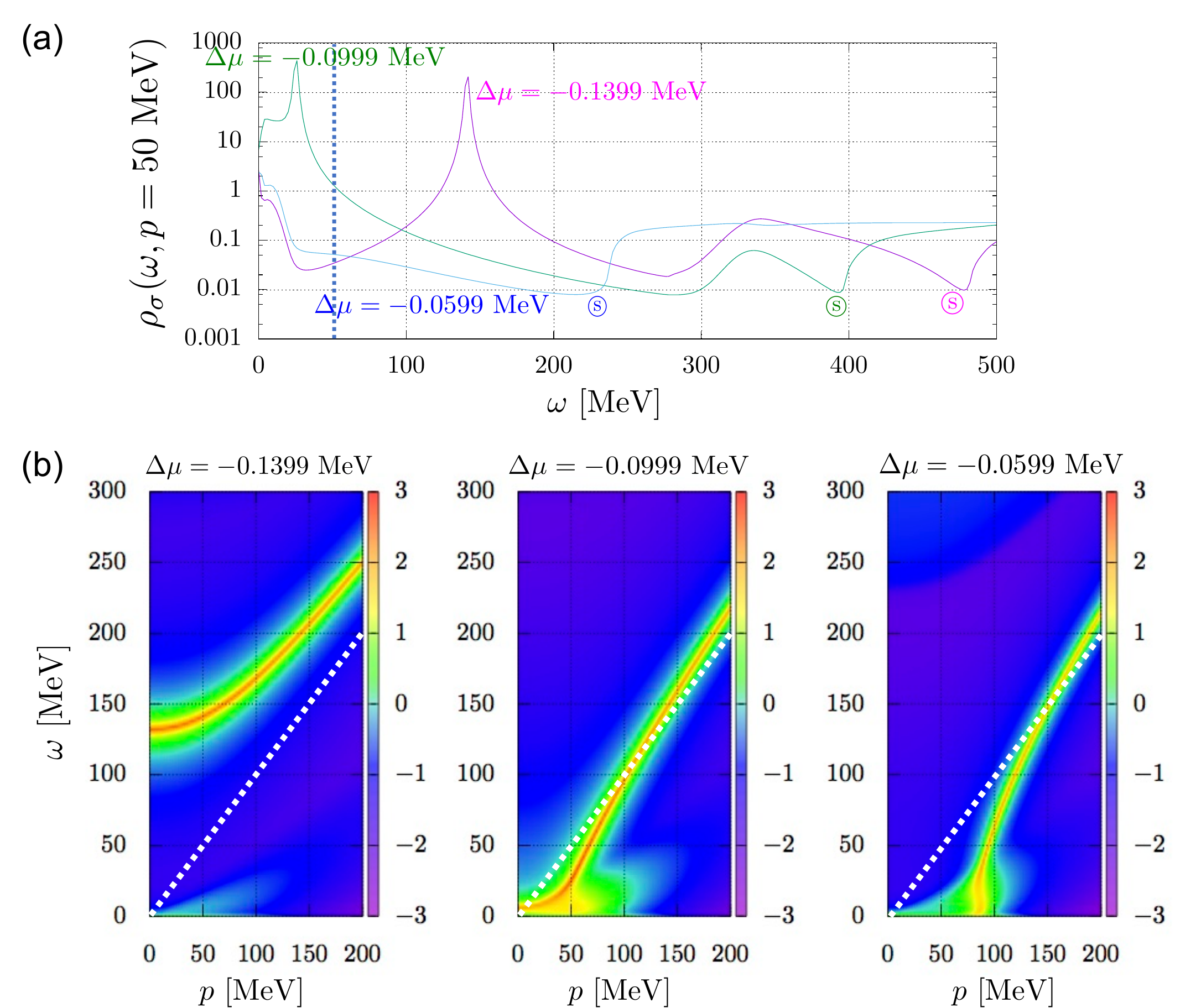}
  \caption{Spectral functions $\rho_\sigma$ in the scalar channel near
   the {\Zcp} critical point for the physical quark mass, $m_q=8.88~\MeV$. The
   temperature is fixed at $T_{c}$
   and the chemical potential
   is increased with an interval  $\Delta\mu = \mu -  \mu_{c}$. (a)
   The spectral functions $\rho_\sigma(\omega,p)$
   at $p=50$ MeV. The dotted vertical line (in blue) shows the energy
   $\omega=p$ on the light cone, while $\textcircled{s}$ indicates
   the threshold energy  of the $2\sigma$ mode.
   (b) The contour maps of $\rho_{\sigma}$ (in logarithmic scale)
   in the $(\omega, p)$-plane
   for the same temperature and the chemical potentials as those in (a).
   The dotted straight line (in white) denotes the light-cone $\omega=p$.
   \label{Fig:c0.00175}}
 \end{figure*}
}
{
 \begin{figure*}[!htb]
  \centering\includegraphics[width=1.6\columnwidth]{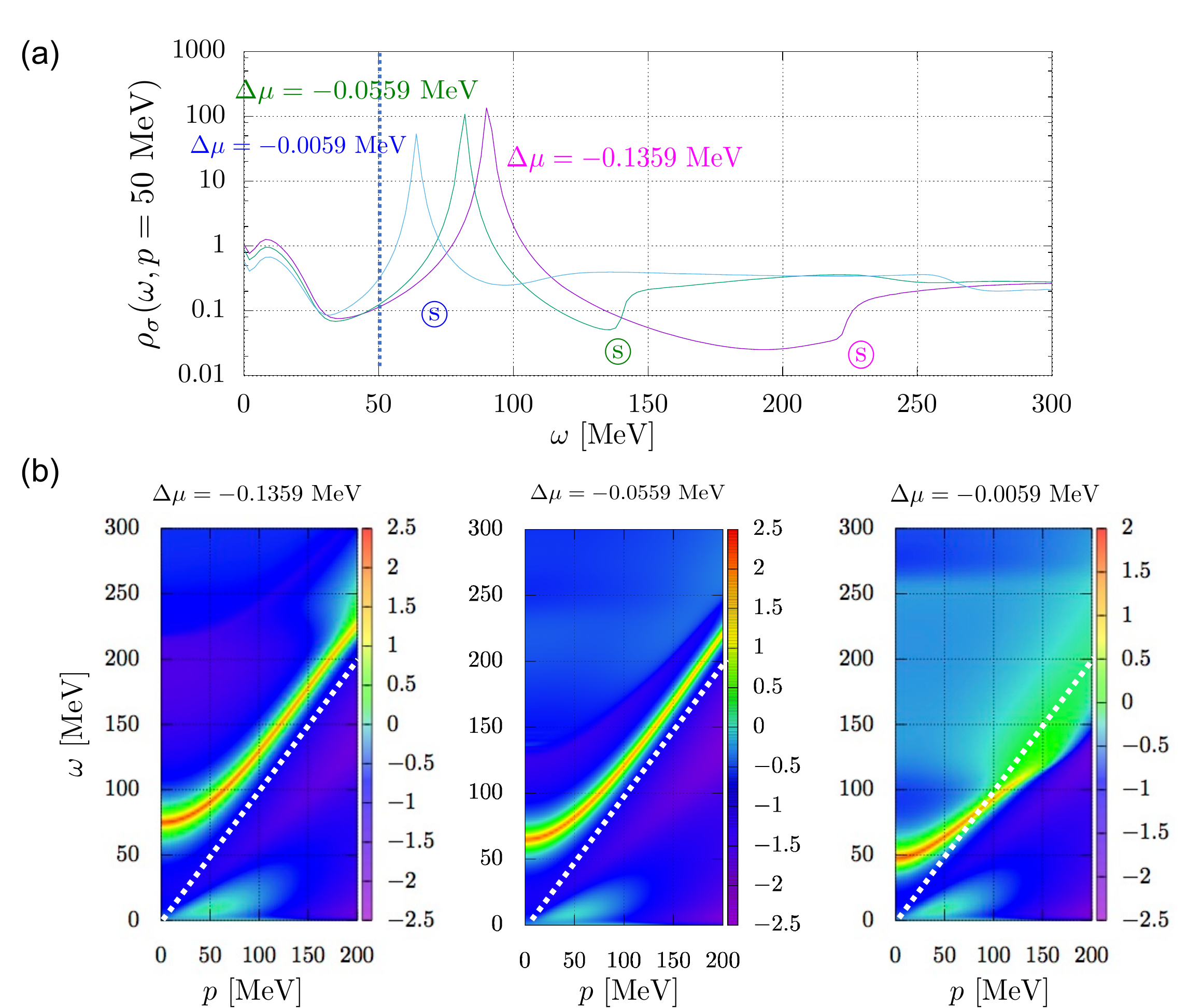}
  \caption{
   The same as Fig.~\ref{Fig:c0.00175}, but for $m_{q}=1.27~\MeV$.
   The temperature is fixed to $T_{c}(m_q)=8.6~\MeV$.
   \label{Fig:c0.00025}}
 \end{figure*}
}

For the sake of the self-containedness, we begin with showing the
numerical result for the physical quark mass $m_q=8.88~\MeV$, which is
essentially the same as that presented in the previous paper
\cite{Yokota:2016tip} but with a slightly better precision.
Figure \ref{Fig:c0.00175}(a) displays the spectral
functions $\rho_\sigma(\omega,p)$ at $p=50$~MeV for a few values of the
chemical potential close to $\mu_c$ at $T=T_c$.
For smaller $\mu$, a peak and bumps are present in the time-like
region, whereas a bump in the space-like region is seen;
the peak in the time-like region represents the $\sigma$ mesonic mode, while
the bump in the space-like region denotes p--h excitations
due to the coupling between the scalar and vector channels
$\langle \bar{\psi}\psi\bar{\psi}\gamma^0\psi\rangle$ at finite $\mu$.
A notable point is  that the $\sigma$ meson peak sitting in the time-like region
goes down to lower energy as
$\mu$ is increased toward $\mu_c$ and eventually
penetrates into the space-like region $\omega < p$ as the system
is quite close to
the critical point. This behavior can be more effectively seen
in the contour $\rho_\sigma(\omega, p)$ mapped onto $(\omega,p)$-plane:
Fig.~\ref{Fig:c0.00175}(b) displays the downward shift of the peak and
the resultant anomalous dispersion relation in which the group velocity of
the $\sigma$ mode exceeds the speed of light.
Admittedly, it might be caused by the possible violation of the
causality owing to the use of
the three-dimensional regulator~\footnote{We thank
J.~Pawlowski for pointing out this possibility},
and thus it would be desirable to check the regulator dependence
\cite{Pawlowski:2015mia,Strodthoff:2016pxx}.
However, it should be noted that the superluminal group velocity seems
to appear only in the limited situations, contrary to what is expected
by the regulator origin.
Another possibility is that a level repulsion
between $\sigma$ and $2\sigma$ modes causes
the appearance of this tachyonic mode as pointed out
in Ref.~\cite{Yokota:2016tip}.
In fact, Fig.~\ref{Fig:c0.00175}(a) shows that the well-defined
$2\sigma$ threshold energy moves down so that the level repulsion
with the $\sigma$ mesonic mode becomes more effective.

\subsubsection{The physical meaning of the appearance of the tachyonic mode:
analysis with varying quark mass}

In the previous work \cite{Yokota:2016tip}, the physical significance and the
origin of the appearance of the tachyonic mode were, unfortunately, not pursued
but left for a future work. Now we deal with this task.
First of all, we notice that the  $\sigma$-to-2$\sigma$
coupling gets to exist due to the explicit violation of chiral symmetry
dictated by the current quark mass $m_q$. Therefore it should be
intriguing to examine the behavior of the spectral function
for smaller $m_q$ leading to a suppressed $\sigma$-to-2$\sigma$ coupling.

Figure~\ref{Fig:c0.00025} shows the case of a finite but tiny quark mass,
$m_q=1.27$ MeV.
Both figures show a downward shift of the peak of the
sigma mesonic mode and accordingly of the 2$\sigma$ threshold energy,
while
a bump due to the p--h
excitations is clearly seen in the space-like region
as in the case of $m_q=8.88~\MeV$.
Nevertheless, no anomalous behavior with a superluminal group velocity
appears in the dispersion relation.
{
 \begin{figure*}[!htb]
  \centering\includegraphics[width=1.7\columnwidth]{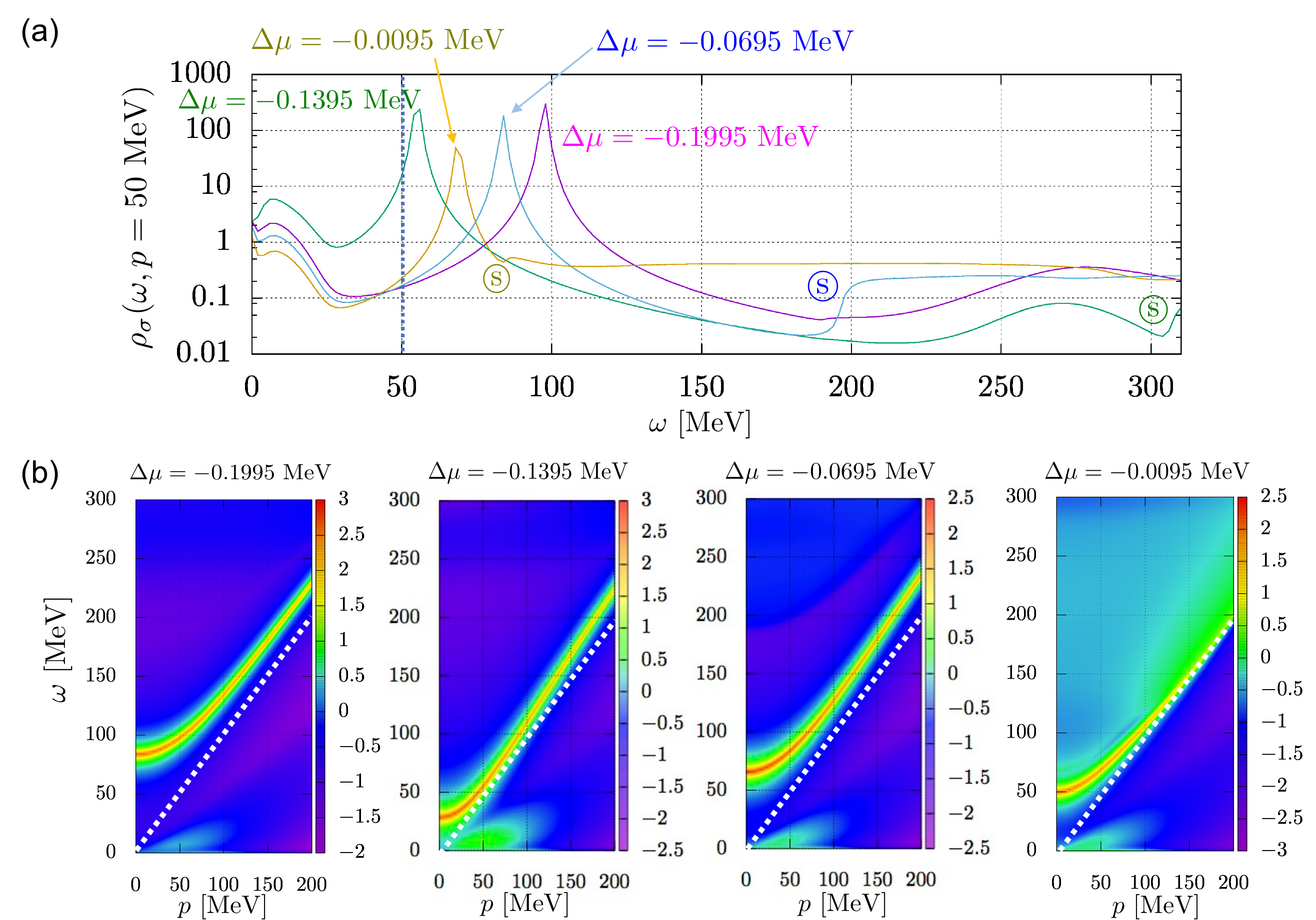}
  \caption{
   The same as Fig.~\ref{Fig:c0.00025}, but for  $m_{q}=3.81~\MeV$.
   The temperature is fixed to $T_{c}(m_q)=6.9~\MeV$.
   \label{Fig:c0.00075}}
 \end{figure*}
}
{\begin{figure*}[!htb]
 \centering\includegraphics[width=1.5\columnwidth]{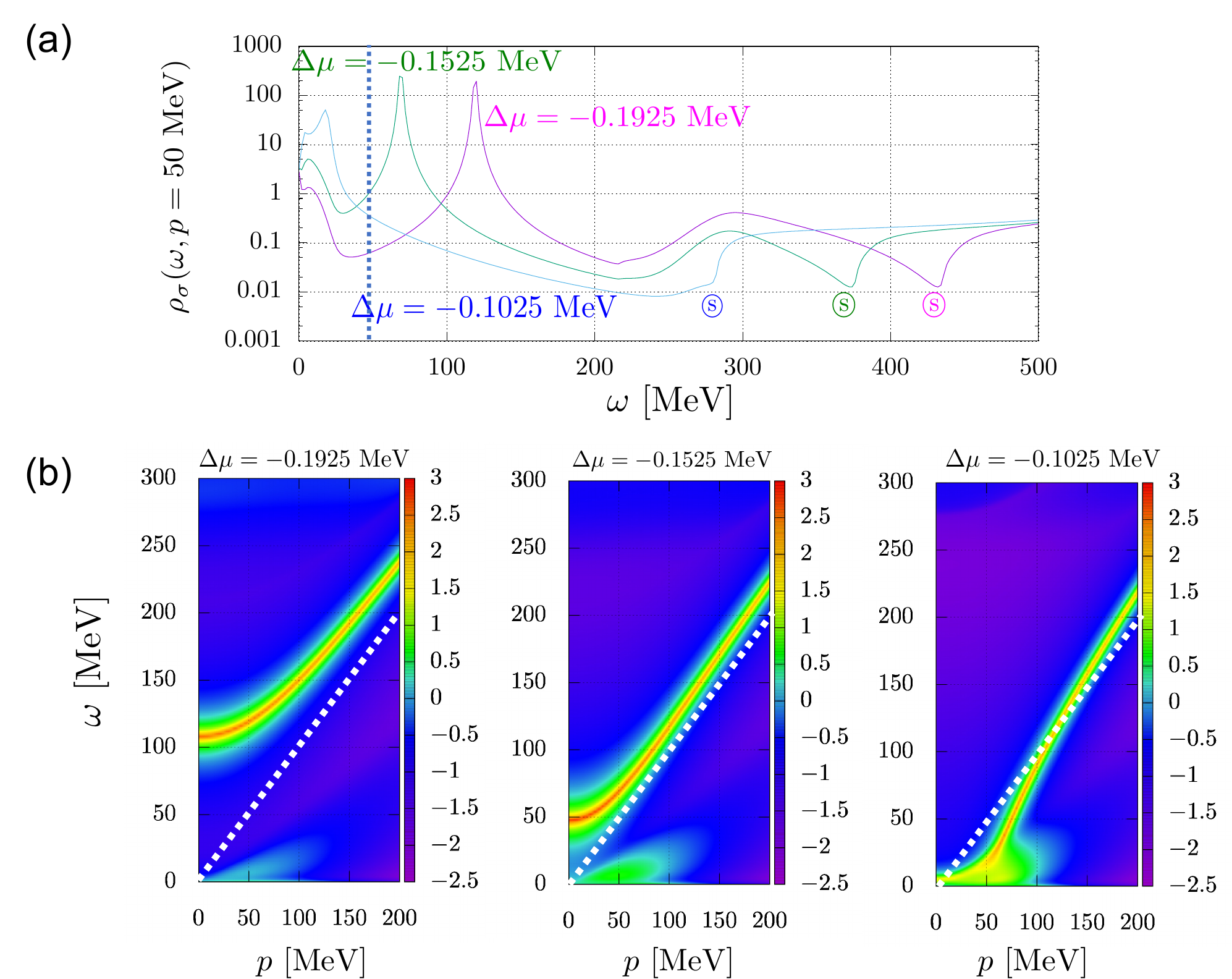}
 \caption{
  The same as Fig.~\ref{Fig:c0.00025}, but for  $m_{q}=5.08~\MeV$.
  The temperature is fixed to $T_{c}(m_q)=6.3~\MeV$.
  \label{Fig:c0.00100}}
 \end{figure*}
}

Figure~\ref{Fig:c0.00075} shows the case when $m_q$ is slightly increased to
$m_q=3.81$ MeV.
Although the anomalous dispersion relation has yet to
appear, one sees a non-monotonic peak shift in the spectral
function. The peak first moves down as in the smaller $m_q$ case
(the first and second panels from the left in (b)),
but turns upward for larger $\mu$
(the third panel), then moves down
again
(the fourth panel) as the system
further approaches the critical point.
Although  such a change should be the result of the competing and
$\mu$-dependent level repulsions
of the  $\sigma$ mesonic mode with the low-lying p--h excitations and
high-lying 2$\sigma$ bump,
it is admittedly not so easy to give a clear account of
such a non-monotonic behavior of the dispersion curve.
Nevertheless, the group velocity of any modes do not become superluminal
for this case.

The situation changes when $m_q=5.08$ MeV as shown in Fig.~\ref{Fig:c0.00100}:
The $\sigma$ peak in the low momentum region is pushed down to the
space-like region. As the $\sigma$ mode in the high-momentum region
remains time-like, the resultant dispersion relation exhibits a superluminal
group velocity
and hence the
anomalous tachyonic behavior appears, as was the case for the physical quark
mass.

\begin{figure}[!ht]
 \centering
 \includegraphics[width=0.5\columnwidth]{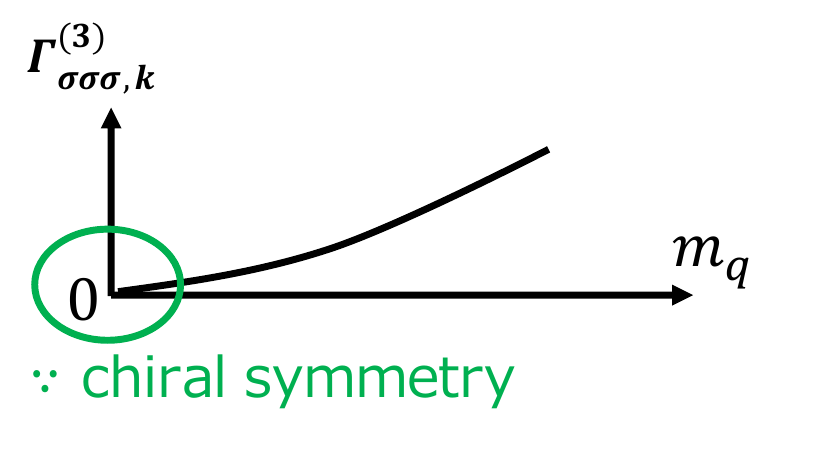}
 \caption{3-point vertex $\Gamma^{(0,3)}_{k,\sigma\sigma\sigma}$ on the
  {\Zcp} critical line as a function of $m_q$.}
 \label{Fig:gamma3}
\end{figure}

\subsection{Discussion}
\label{Sec:discussion}

\subsubsection{Penetration of the sigma mode into space-like momentum region}

In the previous work \cite{Yokota:2016tip}, we pointed out that the
level repulsion between $\sigma$ and $2\sigma$ modes
causes the downward shift of the sigma mesonic mode by demonstrating
that the three-point vertex of sigma $\Gamma^{(0, 3)}_{k, \sigma \sigma
\sigma}$, which gives rise to the level repulsion, strongly affects the
energy shift of the sigma mesonic mode. The downward shift of these
modes at physical quark mass can be found in Fig.~\ref{Fig:c0.00175}(a).

In the chiral limit, $\Gamma^{(0, 3)}_{k, \sigma \sigma \sigma}$
vanishes on the {\Ocp} critical line because the symmetry under the
transformation $\sigma\rightarrow -\sigma$ exists.
Thus the level repulsion owing to
finite $\Gamma^{(0, 3)}_{k, \sigma \sigma \sigma}$
does not happen in the chiral limit.
At finite but small current quark masses,
$\Gamma^{(0, 3)}_{k, \sigma \sigma \sigma}$ on the $\Zcp$ critical line
is expected to slowly increase as $m_q$ increases, as schematically
shown in Fig.~\ref{Fig:gamma3}. Consequently, the level repulsion is
weakened by the small coupling at small current quark masses. This
finding is consistent with the absence of the penetration of the sigma mode
into the space-like momentum region, despite the strong downward shift of
the $2\sigma$ mode.

For larger $m_q$, the group velocity of the sigma mesonic mode
is superluminal for small but finite momenta at $\mu$ close to but below
$\mu_c(m_q)$.
One of the possibilities is that such an appearance of the tachyonic mode
simply shows that the system is unstable, i.e.,
the ground state or the equilibrium state we have assumed is not
the true one in these situations.
A clue may be given by the fact that the tachyonic behavior as well as
the softening itself in the sigma channel occur at finite momenta.
Recently it has been suggested that
the chiral inhomogeneous phase may exist so that it reveals
the first-order phase transition line
between the hadronic phase and the quark--gluon plasma phase
\cite{Nakano:2004cd,Nickel:2009wj,Nickel:2009ke,Muller:2013tya}.
In the present work, we have taken it for granted
that the sigma condensate $\sigma_{0}$ is homogeneous
with vanishing  pion condensate in the equilibrium.
Thus our result might indicate that the system is to undergo a phase
transition to a state with an inhomogeneous sigma condensate near
the $\Zcp$ critical point at $\mu$ smaller than $\mu_c$ at small and
vanishing temperature; as is shown below, any anomalous behavior is not seen
for the excitation modes in the pion channel.

\subsubsection{Soft modes at the critical point}

\begin{figure}[!htb]
 \centering\includegraphics[width=0.95\columnwidth]{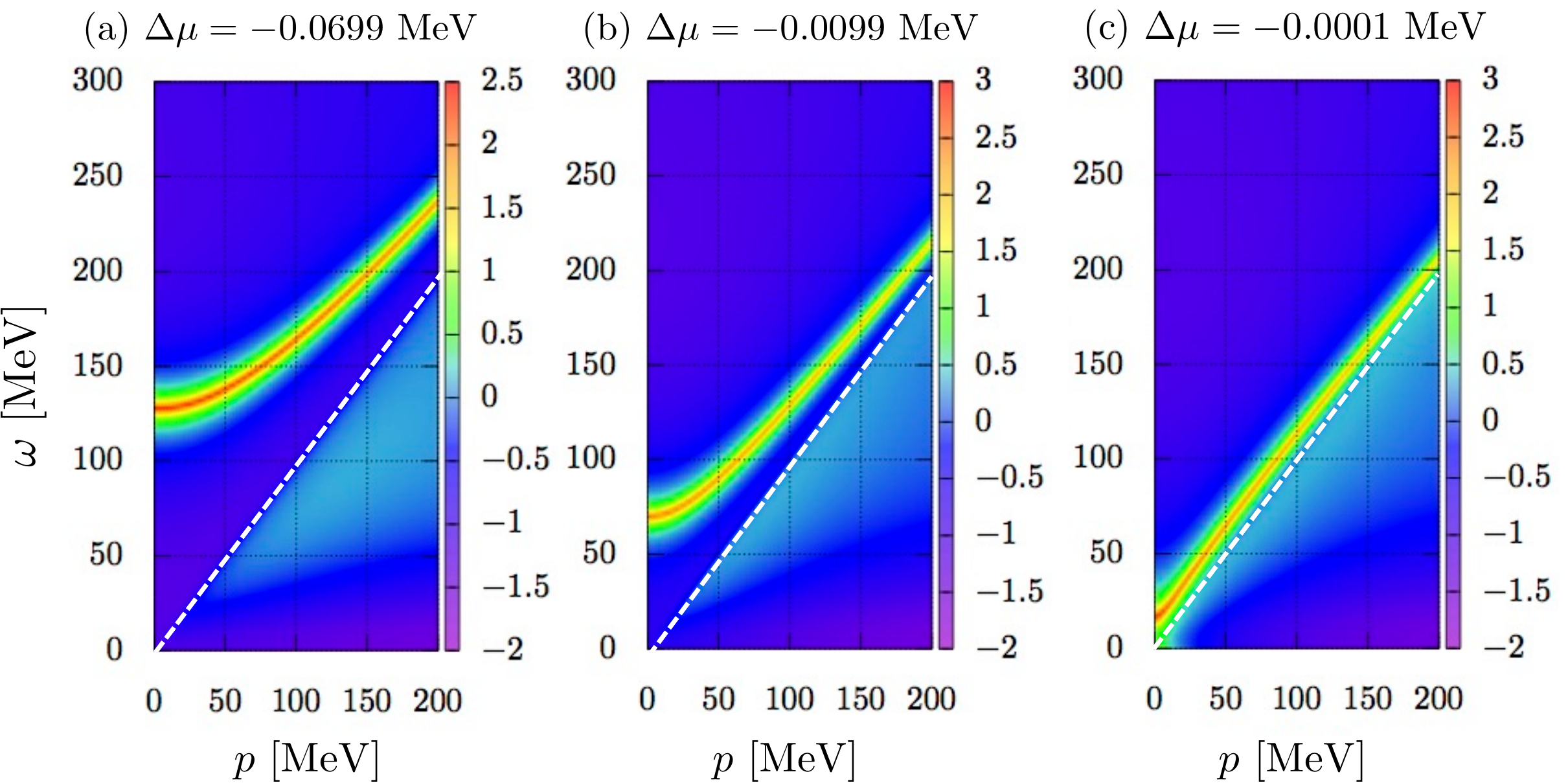}
 \caption{The contour map of $\rho_{\sigma}$ (in logarithmic scale)
  near the $\Ocp$ critical point at $(T, \mu)=(45~\MeV, 260.3599~\MeV)$.
  The temperature is fixed to $45~\MeV$.
  Here $\Delta \mu = \mu - 260.3599~\MeV$ is the relative position
  from the $\Ocp$ critical point.
  The dotted straight line (in white) denotes the light-cone $\omega=p$.
  \label{Fig:O4}}
\end{figure}
\begin{figure}[!htb]
 \centering\includegraphics[width=0.45\columnwidth]{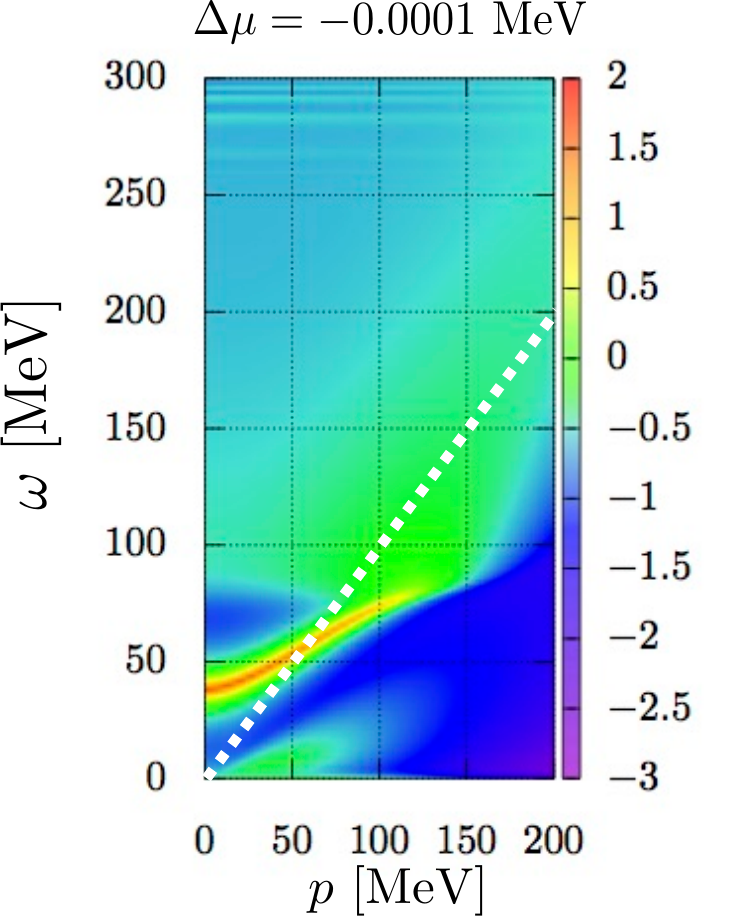}
 \caption{The same as Fig.~\ref{Fig:c0.00025}, but for
 $\Delta \mu = -0.0001~\MeV$.\label{Fig:c000025toCP}
 }
\end{figure}

Before closing this section, we discuss the systematic view of
the soft modes on the $\Ocp$ and $\Zcp$ critical lines
on the basis of the results given above.
Figure \ref{Fig:O4} shows the contour map of $\rho_\sigma$ near the
{\Ocp} critical point $(T, \mu)=(45~\MeV, 260.3599~\MeV)$ on the
{\Ocp} critical line nearby the tricritical point.
One sees a clear ridge in the time-like momentum region ($\omega > p$) as
can be described by a dispersion relation $\omega_{\sigma}(p)$, which is
an isoscalar one-particle mode, i.e., the sigma mesonic mode.
As the system approaches the critical point, $\omega_{\sigma}(p)$ moves
down toward zero energy, strongly in the lower-momentum region and
eventually almost touches the zero energy for vanishing momenta.
Thus one sees that Fig.~\ref{Fig:O4} beautifully shows that
the sigma mesonic mode is the soft mode of the $\Ocp$ critical point.

At finite but small quark mass
$m_q=1.27$~MeV for instance, one finds the downward shift of the peak,
which does not, however, reach $\omega=0$ even at the immediate vicinity
of the critical point, as shown in Fig.~\ref{Fig:c000025toCP}.
Thus the sigma meson is not a soft mode at the $\Zcp$
critical point. This picture is consistent with the arguments of
Ref.~\cite{Fujii:2004jt, Son:2004iv}.
A remark is in order here:
The dispersion curve of the sigma mesonic mode in the right panel of
Fig.~\ref{Fig:c0.00025}(b) and Fig.~\ref{Fig:c000025toCP} penetrates
into the space-like momentum region in the high-momentum region.
This anomalous and would-be interesting behavior is, however,
can be an artifact largely caused by our approximation
in contrast to the tachyonic behavior seen in the low-momentum region for larger
$m_q$.
First of all, we note that since
the $\sigma$ propagator is approximated in terms of
the $k$-dependent screening mass $M_{\sigma}(\sigma_0)$
as given by Eq.~\eqref{Eq:GB},
the threshold of the 2$\sigma$ mode is
also necessarily given by the (physical) screening mass
$m_\sigma=M_{\sigma}(\sigma_0)\vert_{k \to 0}$,
and hence the $2\sigma$ threshold at momentum $p$
does not coincide with
the twice of the $\sigma$-peak energy at momentum $p/2$:
For instance, when $m_q=8.88$~MeV and $\mu$ is not too close to
$\mu_c$, as seen by $\Delta \mu=-0.1399$ MeV in
Fig.~\ref{Fig:c0.00175}(a), the 2$\sigma$ threshold
($\simeq 480$ MeV) is larger than
$2\omega_\sigma^{\text{peak}}(p/2) \simeq 260$~MeV.
This larger $2\sigma$ threshold in the present calculation implies that the
level repulsion between the $\sigma$-2$\sigma$ modes is underestimated.
Thus, the tachyonic behavior seen in the present work should persist in
the exact treatment where a stronger level repulsion can be expected.
This is also the case with $m_q=5.08$~MeV.
However, when $\mu$ is as close to $\mu_c$ as given by
$\Delta \mu = -0.0001~\MeV$,
the screening mass $m_{\sigma}$ almost vanishes and accordingly
the threshold energy of the 2$\sigma$ mode is
greatly underestimated and given by $\omega\sim p$,
which would act to push down $\omega_\sigma(p)$
at the high momentum region owing to the level repulsion.
Then it in turn leads to the penetration of $\omega_{\sigma}(p)$
into the space-like momentum region at high momenta.

When the current quark mass is
further increased, the sigma mesonic mode can move down to $\omega=0$,
but it is accompanied with the appearance of the tachyonic mode, which
may indicate the existence of a true ground state such as inhomogeneous
chiral condensate, as emphasized above.

We also investigate the spectral function in the pion channel.
In the chiral limit, we obtain the dispersion relation of the pion in
the broken phase as $\omega=p$, which is consistent with
the fact that pions become Nambu-Goldstone modes.
At finite current quark mass, the dispersion relation of the pion mode
is found to hardly change near the $\Zcp$ critical points and does not show
any critical behavior, in contrast to the case of the sigma channel.

\section{Summary}\label{Sec:Summary}

We have explored
possible character change of the low-energy modes in
the scalar channels around the QCD critical point with varying the
current quark mass $m_q$ to elucidate the physical meaning of the
anomalous behavior of the sigma mesonic and associated particle--hole
modes found in our previous work \cite{Yokota:2016tip}:
For this purpose,
we have calculated the spectral functions
in the mesonic channels as well as the thermodynamic quantities  using
the functional renormalization group (FRG) method with the 2-flavor
quark--meson model for varied quark masses.

In the first part, we have given the complete and quantitative phase structure
focusing on the change of the nature and the location of the critical point
in the three-dimensional space $(T,\ \mu,\ m_q)$
consisting of temperature $T$, quark chemical potential $\mu$ and $m_q$,
as shown in Fig.~\ref{Fig:CPlocation}:
In the chiral limit, the chiral transition is of second order for vanishing or
relatively smaller $\mu$, and the critical points form an $\Ocp$ critical
line until the tricritical point at which the phase transition turns to
a first order for larger $\mu$.
For a finite $m_q$, the phase change is crossover for vanishing and
small $\mu$ and then the $\Zcp$ critical line extends from the tricritical point.

We have shown the spectral function $\rho_{\sigma}(\omega, p)$ as a function of
$\omega$ for some $p$ and also given the contour map of
$\rho_{\sigma}(\omega, p)$ for whole $(p, \omega)$ plane in the low
energy region extracted from $\rho_{\sigma}(\omega, p)$;
the contour map clearly exhibits the dispersion relations (curves) of the sigma
mesonic mode and particle--hole (phonon) mode as a ridge or bump of the
contour, and thereby clarified how the spectral properties of the
low-energy modes in the sigma channel are changed as $\mu$ approaches
the critical chemical potential from below with $T$ being fixed at $T_c$
for each current quark mass.

At the physical current quark mass $m_{q} = 8.88~\MeV$
reproducing the empirical value of the pion in vacuum,
the dispersion relation $\omega_{\sigma}(p)$ of the sigma mesonic mode
moves down and penetrates into and stay in the space-like region ($\omega\,<\,p$)
in the low-momentum regime slightly {\em before}
the system reaches the critical point:
Such a drastic downward shift of the dispersion relation necessarily
makes the sigma mesonic mode superluminal at finite momenta, 
as was found in the previous paper \cite{Yokota:2016tip}.

Although one might suspect that the existence of such a superluminal mode could be
an artifact due to the violation of the causality owing to the use of
the three-dimensional regulator,
the present analysis tracing the spectral change of the
scalar mode with varying current quark mass has shown
that it has a physical origin. First of all, the downward shift
can be understood in terms of the level repulsion between the sigma
and the two-sigma mode \cite{Yokota:2016tip}
and this  effect is suppressed for small $m_q$ and
vanishes in the chiral limit because the three-point vertex of
the sigma $\Gamma_{k,\sigma\sigma\sigma}^{(0,3)}$ vanishes on the critical line
due to the chiral symmetry.
Our analysis with smaller $m_q$ shows the absence of
the superluminal velocity
for such $m_q$: Indeed at as small as $m_{q}=1.27~\MeV$,
$\omega_{\sigma}(p)$ does not penetrate into the space-like momentum region
in the low-momentum region and no superluminal mode appears,
although $\omega_{\sigma}(p)$ moves down
toward the low-energy region as the system approaches the critical point.
For $m_{q} = 3.81~\MeV$, the dispersion curve of the sigma mesonic
mode shows a non-monotonic behavior
as the system approaches
the critical point, but does not show any superluminal behavior.
For $m_{q} \geq 5.08~\MeV$, the dispersion curve is further pushed down
before it moves up and the downward shift makes the sigma mesonic mode
superluminal as at $m_{q} = 8.88~\MeV$.
These results strongly suggest that the strong downward shift of
$\omega_{\sigma}(p)$ leading to the
appearance of the tachyonic mode with a superluminal velocity has a
definite physical origin due to the $\sigma$-2$\sigma$ coupling caused
by the explicit breaking of chiral symmetry by the current quark mass.
It should be emphasized that the tachyonic dispersion relation appears
when $\mu$ is close to but smaller than  the critical value $\mu_c$.
A natural interpretation of the appearance of the tachyonic mode
at {\em finite momenta} is that the assumed equilibrium state
is unstable against a new state with an inhomogeneous
$\sigma$ condensate
prior to the $\Zcp$ critical point where the phonon mode mainly
composed of particle-hole excitations would overwhelm the $\sigma$
mesonic sector \cite{Fujii:2004jt,Son:2004iv}.

Incidentally,
 on the $\Ocp$ critical line for the chiral limit,
the dispersion relation $\omega_{\sigma}(p)$ of the sigma mesonic mode moves
down toward zero energy at $p=0~\MeV$, which means that the sigma mesonic mode is
the soft mode of the $\Ocp$ critical point.
On the $\Zcp$ critical line at finite but small quark mass,
$m_{q}=1.27~\MeV$ for instance,
$\omega_{\sigma}(p)$ never touch the zero energy near the critical point,
which is in accordance with the fact that the sigma
mesonic mode is not the soft mode of
the $\Zcp$ critical point \cite{Fujii:2004jt,Son:2004iv}.

The present analysis on the excitation modes admittedly
only suggests the possibility that the high-density matter undergoes a
phase transition to an inhomogeneous state. To have a definite answer of
the new state,
it is necessary to develop methods
to deal with  non-uniform equilibrium states
in the framework of FRG. This is a challenging task and left as a future work.

\section*{Acknowledgments}
T.~Y. was supported by the Grants-in-Aid for JSPS fellows
(Grant No. 16J08574).
T.~K. was supported by JSPS KAKENHI Grants (Nos. 16K05350 and 15H03663)
and by the Yukawa International Program for Quark-Hadron Sciences (YIPQS).
K.~M. was supported by JSPS Grant 16K05349,
the Grants-in-Aid for Scientific Research on
Innovative Areas from MEXT (Grant No. 24105008)
and National Science Center, Poland under grants:, Maestro DEC-2013/10/A/ST2/00106.
He also acknowledges support from RIKEN iTHES group.
Numerical computation in this work was carried out at the
Yukawa Institute Computer Facility.

\end{document}